\documentclass[journal,twoside,web]{ieeecolor}
\usepackage{tmi}
\usepackage{cite}
\usepackage{amsmath,amssymb,amsfonts}
\usepackage{algorithmic}
\usepackage{graphicx}
\usepackage{textcomp}
\usepackage[center]{caption}
\def\BibTeX{{\rm B\kern-.05em{\sc i\kern-.025em b}\kern-.08em
    T\kern-.1667em\lower.7ex\hbox{E}\kern-.125emX}}
\begin{document}
\title{CCS-GAN: COVID-19 CT-scan classification with very few positive training images}
\author{Sumeet Menon, Jayalakshmi Mangalagiri, Josh Galita, Michael Morris, Babak Saboury, Yaacov Yesha, Yelena Yesha, Phuong Nguyen, Aryya Gangopadhyay, David Chapman

\thanks{Submitted September $30^{th}$, 2021. This research was supported by NSF award titled {\em RAPID: Deep Learning Models for Early Screening of COVID-19 using CT Images}, award \# 2027628. This work was also supported in part by the NSF IUCRC program as part of the Center for Advanced Real Time Analytics (CARTA) Research Experience for Undergraduates (REU).}
\thanks{ Sumeet Menon, Jayalakshmi Managalagiri, Josh Galita, David Chapman, Aryya Gangopadhyay, Yaacov Yesha and Phuong Nguyen are affiliated with the University of Maryland, Baltimore County (address: 1000 Hilltop Circle, Baltimore, MD, 21250) (email: sumeet1@umbc.edu).  }
\thanks{Yelena Yesha is affliated with the University of Miami}
\thanks{Michael Morris is affiliated with the University of Maryland, Baltimore County, National Institutes of Health Clinical Center and Networking Health ( address: 331 Oak Manor Drive, Suite 201, Glen Burnie MD, 21061). Babak Saboury is affiliated with University of Maryland, Baltimore County and National Institutes of Health Clinical Center (address: 9000 Rockville Pike, Building 10, Room 1C455, Bethesda, MD 21201)}
}

\maketitle

\begin{abstract}

We present a novel algorithm that is able to classify COVID-19 pneumonia from CT Scan slices using a very small sample of training images exhibiting COVID-19 pneumonia in tandem with a larger number of normal images.  This algorithm is able to achieve high classification accuracy using as few as 10 positive training slices (from 10 positive cases), which to the best of our knowledge is one order of magnitude fewer than the next closest published work at the time of writing. Deep learning with extremely small positive training volumes is a very difficult problem and has been an important topic during the COVID-19 pandemic, because for quite some time it was difficult to obtain large volumes of COVID-19 positive images for training.  Algorithms that can learn to screen for diseases using few examples are an important area of research.  We present the Cycle Consistent Segmentation Generative Adversarial Network (CCS-GAN).  CCS-GAN combines style transfer with pulmonary segmentation and relevant transfer learning from negative images in order to create a larger volume of synthetic positive images for the purposes of improving diagnostic classification performance.  The performance of a VGG-19 classifier plus CCS-GAN was trained using a small sample of positive image slices ranging from at most 50 down to as few as 10 COVID-19 positive CT-scan images. CCS-GAN achieves high accuracy with few positive images and thereby greatly reduces the barrier of acquiring large training volumes in order to train a diagnostic classifier for COVID-19.

\end{abstract}



\section{Introduction}

Although deep learning algorithms have achieved high performance in cross validated diagnostic tasks including screening for COVID-19 from X-ray and CT modalities, these results have been obtained overwhelmingly using extensive volumes of training data, including large volumes of COVID-19 positive images obtained from many cases. The use of large volumes of positive cases for training, however, is problematic, especially for a novel disease, as there may be a substantial lag between when the disease becomes a major public health concern and when large training datasets become publicly available, especially given HIPAA and IRB considerations \cite{b22} \cite{b23}.  This has led to a great deal of research in order to determine how a deep learning algorithm can best screen for a disease when few positive training samples are available \cite{b15}, \cite{b16},\cite{b17},\cite{b18}.  Successful deep learning-based COVID-19 diagnostic classification from CT-scans has been demonstrated using hundreds of positive cases for training \cite{b15}, \cite{b16},\cite{b17},\cite{b18}, but it is desirable to demonstrate that accurate classification is possible using even fewer training cases, especially as a preventative measure for a potential future pandemic for which adequate training examples may once again be difficult to obtain.  As such, throughout this paper, an assumption is made that negative (normal) images are prevalent and accessible, whereas positive (COVID-19) images are rare and/or difficult to obtain. Under such an assumption, reasonable metrics for the evaluation of an AI-based screening algorithm including accuracy and AUC obtainable using a given number of positive images, even if larger volumes of negative images may be used freely as necessary for training of a deep neural network in order to achieve adequate classification performance.  To the best of our knowledge, we present the first deep learning based COVID-19 classification algorithm capable of achieving a cross validated test accuracy of 99.00\% (VGG-19) and 98.17\% (AlexNet) while using only 10 COVID-19 positive CT-scan cases for training.  Furthermore, the algorithm presented obtains these results using only a single CT-scan slice per positive training case.  These results are obtained by using a Cycle-Consistent Segmentation-Generative Adversarial Network (CCS-GAN) that is designed to generate high quality COVID-19-infected pulmonary images using as few positive training examples as possible.  CCS-GAN incorporates style transfer based on Cycle-GAN, automated intensity-based pulmonary segmentation, and transfer learning from relevant CT-scan images, all of which greatly reduce the need for positive samples and allow the method to learn from a highly skewed training dataset with many more negative samples.  An ablation study included with this work demonstrates that each of these underlying techniques alone is insufficient to achieve the presented results, and rather all of these techniques must be combined as part of the CCS-GAN methodology in order for high classification accuracy to be possible using very few positive cases for training.

\begin{figure*}
\includegraphics[width=\linewidth]{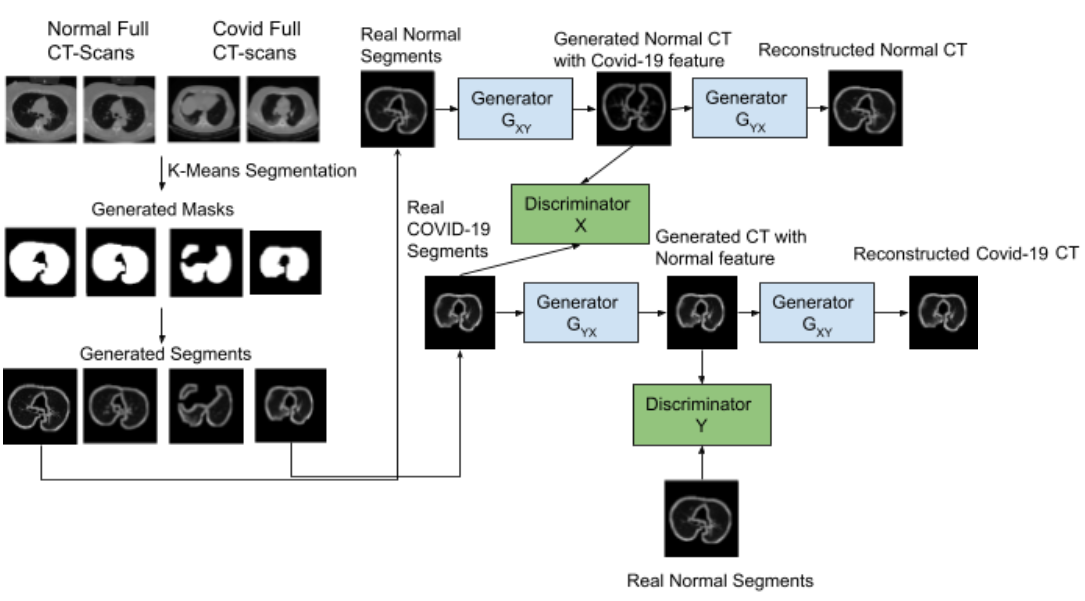}
\caption{CCS-GAN Approach}
\label{fig:CCS-GAN Approach}
\end{figure*}

\section{Related Work}

There have been a number of related works that have investigated the use of GANs to improve the performance of COVID-19 screening from CT scans with reduced training volumes.  However, to the best of our knowledge, all recent studies have made use of hundreds of positive cases for training \cite{b15}, \cite{b16}, \cite{b17}, \cite{b19}, \cite{b20}.  Hundreds of cases, although small by deep learning standards, is still a substantial training volume to obtain during a pandemic, and the intent of CCS-GAN is to determine how advanced methods may be able to greatly reduce the number of positive images needed for potential future events.  Loey et al. introduced the use of conditional GANs (cGAN) for the generation of deep synthetic COVID-19  CT  scans  \cite{b15}. Their cGAN methodology generates both normal and COVID-19 images by conditioning using the category label (COVID/non-COVID).  Goel  et  al.  developed a similar  approach making use of the InceptionV3 network with whale optimization for hyperparameter  tuning  \cite{b16}.  Li et al \cite{b19} extended these approaches by combining GANs with ensemble learning and attention mechanisms. Mangalagiri et al. also proposed an algorithm for generating 3D diagnostic quality COVID-19 CT scans with a conditional GAN architecture \cite{b20}. This method mainly focused on generating the entire CT volume through subdivision into blocks and focusing on block-wise synthesis rather than slice-wise synthesis.  All of these methods have demonstrated improved performance of a binary classifier with a limited number of positive cases for training.  However, all of these methods still require hundreds of positive cases or more to achieve their reported classification accuracies over a withheld testing set.  We are unaware of any works prior to CCS-GAN that have been able to demonstrate comparable diagnostic classification accuracy using on the order of 10 positive cases for training.
Several recent works that have looked at the use of style transfer as a foundation for deep fake CT image synthesis.  Similar to CCS-GAN, most of these approaches have made use of the cycle-consistency loss approach from CycleGAN as a backbone approach.  However, none of these works have combined this approach with automated pulmonary segmentation, and as such, these approaches may be susceptible to artifacts in regions unrelated to COVID-19 disease presentation \cite{b10}, \cite{b21}, \cite{b7}, \cite{b18}.  Sandfort et al. show that Cycle-GAN can be used to generate synthetic CT images by learning the transformation of contrast to non-contrast CT \cite{b10}. Ghassemi et al. \cite{b21} also show the use of Cycle GANs for improved COVID-19 classification with transfer learning, but they have used a total of 3163 images, which is far more than the proposed approach. Jin et al. incorporates prior training on a similar dataset in order to regularize the model and improve performance, especially with limited training volumes \cite{b7}.   Liu   et   al.   \cite{b18}  generates  full  CT scan  volumes  using  feature  in-painting to  insert  COVID-19 opacities via  alpha-opacity  blending.  Transferring disease presentation as a style is an active area of research, but more work is necessary to demonstrate that style-transfer is feasible for the CT-scan modality using a small number of positive images, as well as to prevent style-transfer from introducing artifacts in background regions that are not affected by the target disease.

Very few works have looked at the possibility of using pulmonary segmentation in order to greatly simplify the problem of generating high quality CT-scan slices of COVID-19 pneumonia, and instead most have focused on attempting to generate the full CT-scan slices including irrelevant anatomy.  We demonstrate that pulmonary segmentation, however, can have a major positive impact on the performance of GAN techniques, especially when training with few examples, because when training data volumes are small, the GAN could be distracted by learning to generate bones, organs, and other non-lung anatomy which are irrelevant to COVID-19 at the expense of decreased ability to learn relevant pulmonary features.  
Although Jiang et al. \cite{b17} was the first to make use of pulmonary segmentation to generate synthetic  COVID-19 CT  scan  slices  using  a  dual  generator/discriminator with dynamic element-wise sum, their segmentation approach requires annotation  of  the  infected  region  of  the  lung  which  may be tedious for a radiologist to annotate for training data purposes \cite{b17}.  We are unaware of a comparable GAN approach that takes advantage of automated pulmonary segmentation without the need for pixelwise annotated training data.

Although there have been very few works to take advantage of pulmonary segmentation for the purpose of deep fake COVID-19 synthesis, there are several recent works that have explored segmentation toward the identification of lesions.  However, we do not believe these works to be directly relevant to the proposed research aims because lesion segmentation (CADe) is a very different task from diagnostic classification (CADx).  Deep Convolutional GAN (DCGAN) \cite{b9} and Conditional-GAN \cite{b11} were used to augment medical CT images of liver lesions and mammograms, yielding improved CNN based classification accuracy of malignancy \cite{b12}, \cite{b13}.  
Z. Xu et al. proposed GASNet in \cite{b6} which is a 3D segmentation framework containing a segmentation network with an embedded GAN to segment the pixel boundaries of COVID-19 opacities in CT-scans.  Although substantial progress has been made for the purposes of lesion segmentation, more work is necessary to determine the extent to which segmentation can improve the performance of GAN based 
deep fake image synthesis at low training volumes.

\section{Methods}

\subsection{CCS-GAN}
Figure \ref{fig:CCS-GAN Approach} describes the CCS-GAN approach.  The input dataset is defined as a tuple ($X_N$ , $X_C$) where $X_N$ is the set of normal images, and $X_C$ is the set of images exhibiting COVID-19 pneumonia infection.  Both the normal and covid infected images are segmented using binary K-means / OTSU thresholding to extract the pulmonary regions by creation of a binary segmentation mask.  This segmentation mask can be extracted due to the large intensity difference in radiodensities between lung (less dense) and tissue (denser), even when the lung region is potentially affected by ground glass opacities due to pneumonia.  
After this, the generator models are pre-trained using a cycle-consistent pre-training procedure in which the normal images are employed for both the $X$ and $Y$ categories.  Subsequently, the cycle GAN component is trained using unmatched pairs of normal images $X_N$ and COVID-19 images $X_C$.  Finally, the dataset for use with the classifier is augmented through the generation of deep-fake COVID-19 infected images 
through style transfer from additional normal images.  This multi-faceted approach minimizes the number of positive real cases necessary for training either the GAN or the classifier and allows the methodology to achieve high classification accuracy in the presence of extremely high class imbalance between positive/negative examples. The training set only exhibits between 10 and 50 positive images, with up to 2000 normal images.

\subsection{Intensity-Based Pulmonary Segmentation}
Intensity-based pulmonary segmentation is an effective training-free way to extract the lung region.  Intensity-based segmentation is effective due to the large differences in radiodensity between the lung region, which is mostly air, and the denser surrounding tissue.   


In order to extract the lung region for purposes of pulmonary segmentation, binary K-means thresholding was performed.  K-means attempts to minimize the intra-cluster variance as follows,

\begin{align}
J =&     \operatornamewithlimits{\sum}_{j = 1 }^{k } \operatornamewithlimits{\sum}_{i = 1 }^{n} || x_i^j - c_j ||^2
\end{align}

In this equation, $J$ is the objective function, $k$ is the number of clusters, $n$ is the number of cases, $x_i^j$ is the $i^th$ case,  $c_j$ is the centroid for cluster j,  and $|| x_i^{(j)} - c_j ||^2$ is the distance function.  The special case of k-means clustering with only two cluster centers is mathematically equivalent to the Otsu thresholding technique for binary image segmentation, which minimizes the intra-class variance between pixel intensity histograms over the image as follows,

\begin{align}
\sigma_w^2(t) =&  q_1(t)\sigma_{1}^{2}(t) + q_2(t)\sigma_{2}^{2}(t) 
\end{align}

Where $\mu$  is the mean of the pixel,  is the standard deviation, $q$ is the sum of the probabilities $t$ is the threshold ranging from the minimum value of the pixel to the highest value of pixels. 

\begin{figure}
\includegraphics[width=\linewidth]{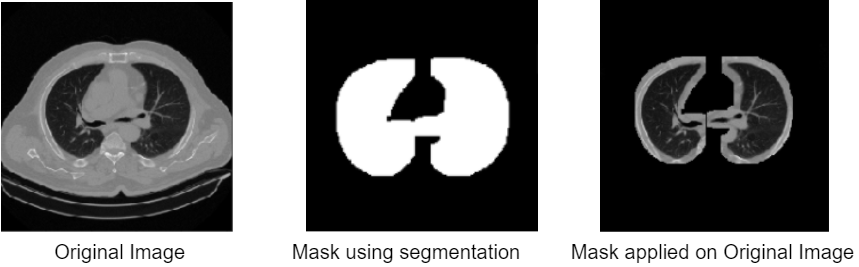}
\caption{Pulmunonary Segmentation}
\label{fig:images}
\end{figure}

Subsequent to thresholding, a fixed series of erosion and dilation steps are performed to suppress noise particularly over blood vessels in the lung.  Air outside the patient is also removed, and the lung mask is identified due to its central positioning within the image frame.    All pixels outside of the lung region are zeroed out to prevent non-lung anatomy tissue from contaminating the GAN and classifier training.

\subsection{Adversarial and Cycle Consistent Loss}

CCS-GAN makes use of combined adversarial and cycle-consistent loss functions as first introduced by cycle-GAN \cite{b3}.  The purpose of the cycle-consistent loss is that one should be able to apply style transfer from normal to COVID-19 and vice versa.  This is accomplished by having two generators $G$ and $F$, and two discriminators $D_X$  and $D_Y$.  Each generator and discriminator makes use of the min/max loss as proposed by Goodfellow as follows \cite{b3}, \cite{b24},

\begin{align}
L_{GAN}{(G,D_Y,X,Y)} &= E_{Y \xrightarrow[]{}{Pdata(Y)}}[logD_Y(Y)] \nonumber \\ &+  
E_{X \xrightarrow[]{}{Pdata(X)}}[1-logD_YG(X)]
\end{align}

\begin{align}
L_{GAN}{(F,D_X,Y,X)} &= E_{X \xrightarrow[]{}{Pdata(X)}}[logD_X(X)] \nonumber \\ &+  
E_{Y \xrightarrow[]{}{Pdata(Y)}}[1-logD_XG(Y)]
\end{align}

The cycle consistency loss ensures that if both generators are applied in a row, then the resulting image should be indistinguishable from the original image.  I.e. for any images: $x \in X$, and $y \in Y$, that $F(G(x)) \approx x$, and that $G(F(y)) \approx y$ as follows \cite{b3},

\begin{align}
L_{cyc}{(G,F)} &= E_{X \xrightarrow[]{}{Pdata(X)}}|F(G(X))-X| \nonumber \\ &+  
E_{Y \xrightarrow[]{}{Pdata(Y)}}|G(F(Y))-Y|
\end{align}

The overall loss function accounts for all the losses is the sum of the constituent adversarial and cycle consistent loss functions as follows,

\begin{align}
L_{GAN}{(G,F,D_X,D_Y)} &= L_{GAN}{(G,D_Y,X,Y)}  \nonumber \\ &+ L_{GAN}{(F,D_X,Y,X)} \nonumber \\  &+  L_{cyc}{(G,F)}
\end{align}

\subsection{Generator Architecture}

\begin{figure*}
\includegraphics[width=\linewidth]{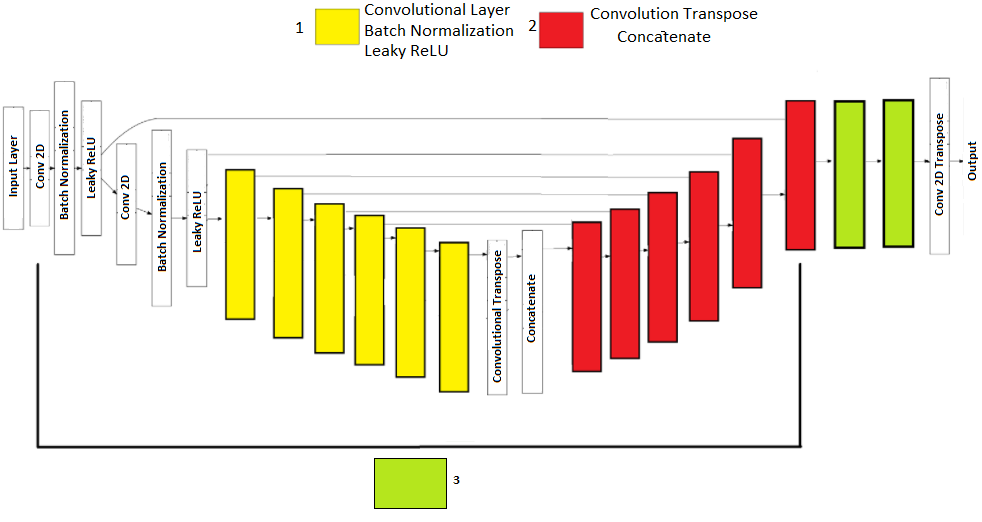}
\caption{Generator Architecture}
\label{fig:Generator_Architecture}
\end{figure*}

\begin{figure*}
\includegraphics[width=\linewidth]{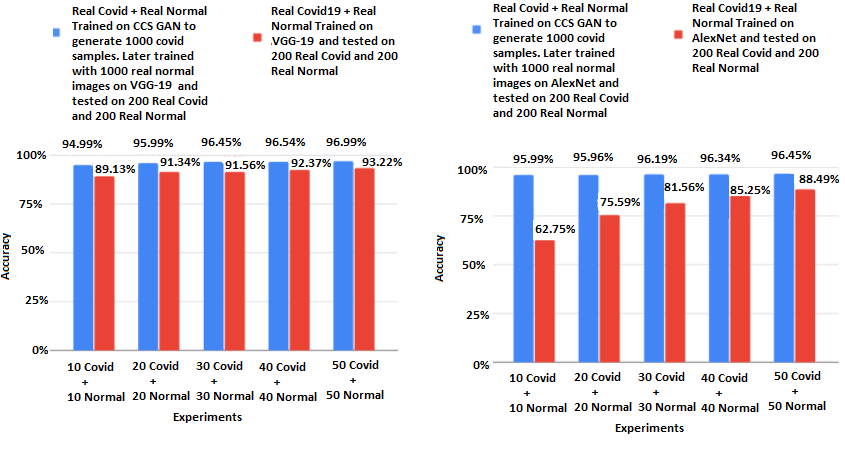}
\caption{AlexNet and VGG-19 Stress Test}
\label{fig:Stress_Test}
\end{figure*}

Figure \ref{fig:Generator_Architecture} describes the generator architecture of CCS-GAN, which is inspired by U-Net \cite{b5}. This architecture is arranged as downsampling blocks (yellow), upsampling blocks (red), and finalizing blocks (green).  The first layer is a conv2d layer which accepts as input a CT-scan slice of size (256,256,1) and outputs features of shape (128,128,1). The next layer is a batch normalization layer, followed by a Leaky ReLU activation. These three layers comprise each of the downsampling blocks as seen in Figure \ref{fig:Generator_Architecture} section 1 (yellow). Furthermore, the Leaky ReLU activation layer has 2 output branches: a downsampling connection and a skip connection.  The skip connection connects the output of a downsampling block of section 1 (yellow) directly to the input of an upsampling block of section 2 (red).  Each upsampling block is comprised of a series of conv2d, transpose, and concatenation layers.  Section 3 (green) exhibits finalizing blocks of alternating conv2D and transpose layers in order to generate a synthetic output image of size (256,256,1).

\begin{figure*}
\includegraphics[width=\linewidth]{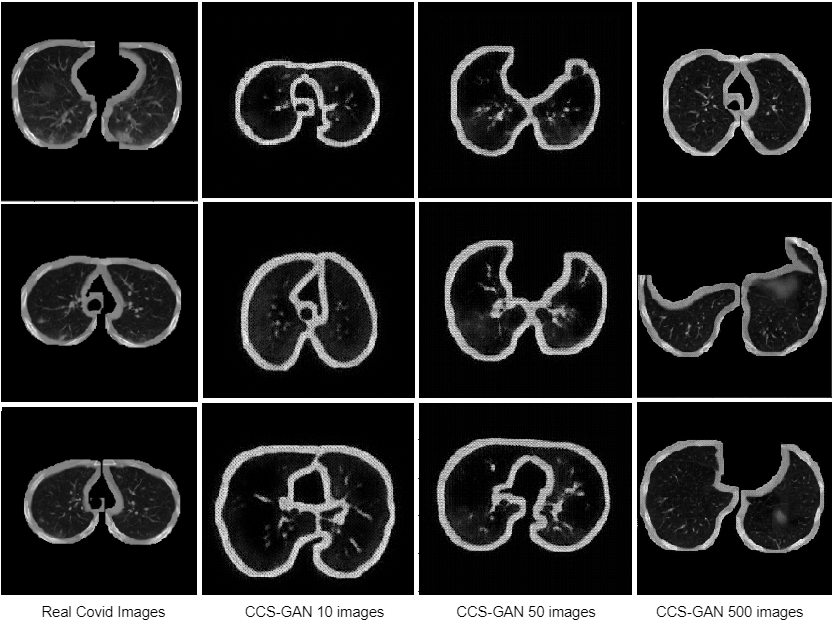}
\caption{Results}
\label{fig:Results}
\end{figure*}

Figure \ref{fig:Results} compares real COVID-19 images versus deep synthetic images as generated by CCS-GAN as trained using a varying number of positive cases.  Left shows real positive cases.  Center left shows generated images as trained with 10 positive cases.  Center right shows generated images as trained with 50 positive cases.  Right shows generated images as trained with 500 positive cases.  As expected, a general improvement in image quality is observed as the number of COVID-19 positive examples is increased, showing detail of blood vessels.  Nevertheless, many pulmonary features are observable using very few positive cases, including those images when CCS-GAN is trained using 50 cases or even 10 cases.

\section{Experimental Design}

CCS-GAN was evaluated both quantitatively and qualitatively, with the overarching goal to determine the extent to which CCS-GAN can improve the ability of a classifier to identify COVID-19 infection using as few positive training images as possible.  Furthermore, a stress test is included in which the number of positive images is reduced to as low as 10.  Two CNN models were used for evaluation of diagnostic classification: AlexNet and VGG-19. These classifiers were trained for 50 epochs from scratch for every experiment. The Adam optimizer was used with a learning rate of $10^{-5}$.  Additionally, another experiment was performed using a form of transfer learning from the normal images only.  For this transfer learning experiment, CCS-GAN was initially pre-trained using 500 random unmatched samples of normal images in which both cycle-consistent classes $X$ and $Y$ consisted of normal CT-slices.  The pre-trained model was subsequently fine-tuned with 1000 normal images and 10 positive COVID-19 images. 
Finally, an ablation study was performed to compare the results produced by the proposed CCS-GAN method versus a baseline GAN.  Throughout this ablation study, individual features of the CCS-GAN are disabled including the cycle-consistent training, the pulmonary segmentation, the cycle-consistent transfer learning with unmatched normal pairs, as well as the U-Net inspired generator architecture (As replaced by ResNet-50).  The key finding of this ablation study is that the entirety of the CCS-GAN approach is necessary to obtain the high quality results, and that if any of these techniques is disabled, the image quality dramatically suffers and becomes unsuitable for the intended use cases.  As such, all of the underlying techniques of CCS-GAN are necessary to achieve the reported performance.

\subsection{Dataset}

\begin{figure}
\includegraphics[width=\linewidth]{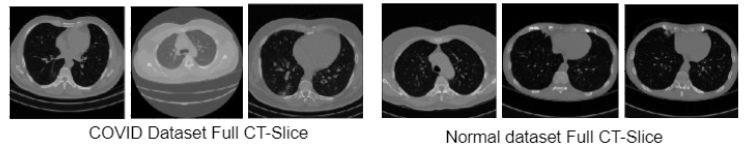}
\caption{Normal Full CT Slices}
\label{fig:images1}
\end{figure}

\begin{figure}
\includegraphics[width=\linewidth]{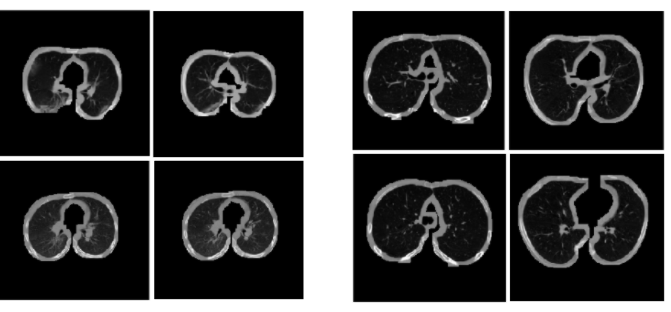}
\caption{Segments of full Slices} 
\label{fig:images2}
\end{figure}

For training and evaluation of the GAN, a  dataset was provided by the Networking Health nonprofit organization consisting of deidentified imagery from 944 CT  scans  from  patients  with  COVID-19 as collected across a diverse set of hospital institutions. However, only between 10 and 50 positive slices were employed for training the GAN in almost all experiments, the only exception being a control experiment in which 500 positive slices were used to observe how this larger number of cases affects qualitative image quality. From these CT scans, only the axial slices in the scan which exhibit substantial lung volume were considered, as determined by the pixel area of the automated lung segmentation mask. The normal CT scans were extracted from the Kaggle  lung  cancer  dataset. This dataset consists of 1,020 normal CT scans. All the slices of both the classes were resized to 256x256. The training dataset for the CCS-GAN consists of COVID-19 and normal CT-slices ranging from 10 to 50 COVID-19 images for each experiment and the testing dataset consisted of 200 images of each class.

\subsection{Quantitative Analysis}
A quantitative analysis was performed to determine the extent to which CCS-GAN can improve the performance of a binary classifier with very few COVID-19 training images. To test the effectiveness, the classifier with CCS-GAN augmented images was compared with a baseline. The backbone classifiers used for this task were AlexNet and VGG-19.  The baseline classifier was trained on a balanced dataset of normal and COVID-19 CT slices ranging from 50 images to 10 images per class. This baseline was compared against a classifier with additional normal images as well as augmented with synthetic COVID-19 images as generated from CCS-GAN.  CCS-GAN was trained using the exact same COVID-19 images as were available to the classifier.  The reported performance metrics are \textit{accuracy} and Area Under the Curve \textit{AUC}.

Figure \ref{fig:Stress_Test} shows a stress test which compares the accuracy of AlexNet and VGG-19 baselines versus the addition of synthetic COVID-19 images from CCS-GAN without transfer learning.  In this stress test, AlexNet and VGG-19 were trained with a small sample of COVID-19 images ranging from 10 images up to 50 images.  In all cases, the addition of CCS-GAN augmented images improves performance relative to the baseline.  In both cases, the most dramatic improvements are obtained when only 10 COVID-19 images were used.  Furthermore AlexNet + CCS-GAN presents a more dramatic improvement in the classifiaction accuracy relative to VGG-19 + CCS-GAN.  Baseline AlexNet achieves an accuracy 62.75\% with 10 COVID-19 images, but this improves to 95.99\% with the addition of CCS-GAN.  
The accuracy of baseline VGG-19 is 89.13\% which improves to 94.99\% with the use of synthetic images as generated by CCS-GAN.  Slightly higher accuracies, with somewhat less substantial improvements, are obtained when the models have access to a larger sample size of COVID-19 images.  With 50 COVID-19 images, baseline AlexNet achieves accuracy of 88.49\%, whereas the inclusion of CCS-GAN improves accuracy to 96.45\%.  Moreover, baseline VGG-19 achieves accuracy of 93.22\% whereas enhancement with CCS-GAN improves accuracy to 96.99\%.  As such, we observe that CCS-GAN greatly improves classification accuracy, and this improvement is particularly pronounced when using only 10 COVID-19 training images relative to 50 COVID-19 images.

Table \ref{accuracy1} shows a final experiment that was performed in which the CCS-GAN was pre-trained using 1000 normal images (split into two groups of 500 images each), prior to fine tuning on 10 COVID-19 CT-scan images and 1000 normal CT-scan images to generate 990 synthetic COVID-19 images for the purposes of training the classifier.  The classifier was then trained using 1000 normal images and 1000 COVID-19 images (990 generated). In this configuration, the classifier outperforms all of the prior experiments and yields an accuracy of 98.17\% and AUC of 0.9989 using AlexNet, as well as an accuracy of 99.00\% and AUC of 0.9995 using VGG19.


\begin{table}[htbp]
\caption{ Comparative performance of classification using CCS GAN with 10 real COVID-19 images.}
\begin{center}
\begin{tabular}{|p{0.9cm}|p{0.5cm}|p{1cm}|p{2cm}|p{2.2cm}|}
\hline
\textit{\textit{Model}} & \textit{\textit{ Test Data CT-slices}} & \textit{\textit{10 Real COVID-19 CT slices + 10 Real Normal CT slices}} & \textit{\textit{1000 Generated COVID-19 CT-slices by CCS-GAN using 10 real COVID-19 CT-slices + 1000 Real Normal CT-slices}} & \textit{\textit{Transfer Learning
1000 Generated COVID-19 CT-slices by CCS-GAN using 10 real COVID-19 CT-slices + 1000 Real Normal CT-slices
}} \\
\hline
\textit{\textit{AlexNet}} &\textbf{200 covid + 200 Normal }  &\textbf{89.13\%} (AUC: 0.9237) &\textbf{94.99\%} (AUC: 0.98753) &\textbf{98.17\%} (AUC: 0.9989) \\
\hline
\textit{\textit{VGG-19}} &\textbf{200 covid + 200 Normal }  &\textbf{62.75\%} (AUC: 0.7284) &\textbf{95.99\%} (AUC: 0.9913) &\textbf{99.00\%} (AUC: 0.9995) \\
\hline
\end{tabular}
\label{accuracy1}
\end{center}
\end{table}

\begin{figure*}
\centering
\includegraphics[width=150mm,height=200mm]{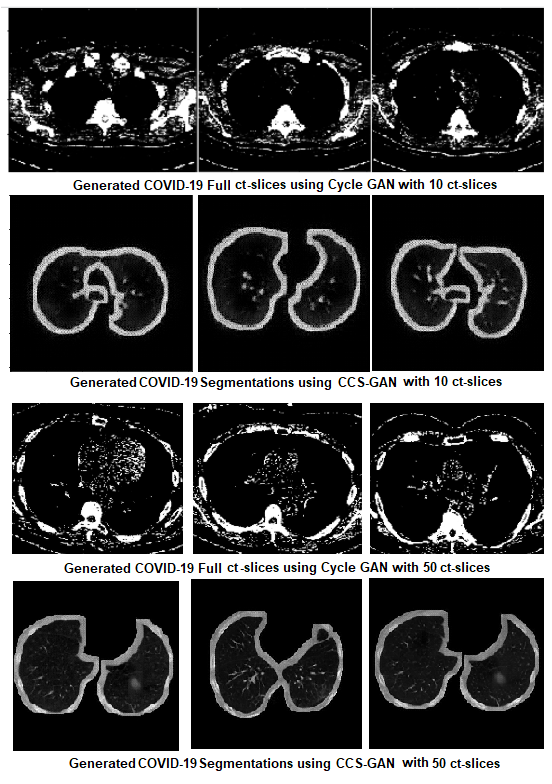}
\caption{Ablation Study}
\label{fig: Ablation Study}
\end{figure*}

\subsection{Ablation Study}

Ablation experiments were performed to determine if the entire CCS-GAN methodology is necessary to obtain the reported results.  In particular, we wish to determine if both the pulmonary segmentation as well as cycle consistent adversarial loss techniques are necessary, or if a simplified version of CCS-GAN using only one of these methods would be adequate for the intended purposes.  The first set of ablation experiments were to generate synthetic COVID-19 images using only pulmonary segmentation, only cycle consistent adversarial loss, with either the U-net like proposed generator or using a ResNet 50 generator architecture.  
The qualitative outcome of these experiments is presented and all of which yield very poor image quality relative to the full CCS-GAN which combines all of these techniques into a single methodology.  

\begin{figure}
\centering
\includegraphics[width=\linewidth]{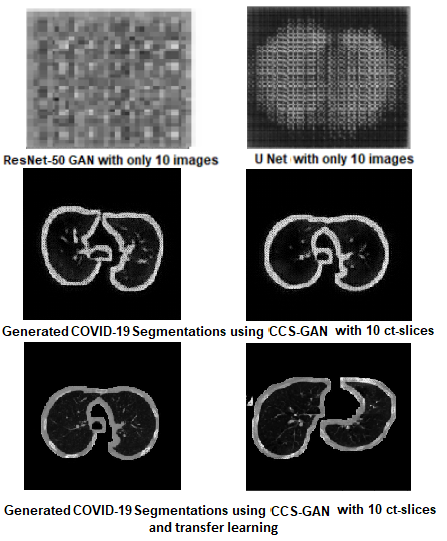}
\caption{GANs vs CCS GAN for 10 images}
\label{fig: Ablation Study GANs}
\end{figure}

Figure \ref{fig: Ablation Study GANs} shows the image quality obtained if cycle-consistent adversarial loss is disabled.  As such, the pulmonary segmentation is used in both cases, but the generator is only trained to generate COVID-19 images using the 10 COVID-19 input image slices.  We also made use of two generator models, the U-net like architecture as used in CCS-GAN, as well as a ResNet50 architecture which is a commonly employed alternative in related literature.  As we see in Figure \ref{fig: Ablation Study GANs}, when a regular GAN is trained using pulmonary segmentation, in both cases the resulting images are unreasonable and do not show any meaningful anatomic structure.  The U-net generator is able to produce the rough silhouette of a lung region, but the structure within this silhouette appears more similar to woven fabric than to pulmonary anatomy.  The ResNet50 generator produces red noise and is unable to even construct the pulmonary silhouette at such low volumes.  By comparison, the CCS-GAN is able to produce qualitatively reasonable results showing pulmonary features.

Figure \ref{fig: Ablation Study} shows the qualitative results of training Cycle-GAN (without pulmonary segmentation) versus CCS-GAN with two training configurations.  Configuration 1 is with a dataset of 10 slices of the COVID-19 class and 10 slices of the normal class, and Configuration 2 is with a dataset of 50 slices of the COVID-19 class and 50 slices of the normal class.  All images are displayed with a standard grayscale window of [-1000 to 1000] HU.  We observe that without pulmonary segmentation, baseline Cycle-Gan does not generate qualitatively reasonable images using such a small COVID-19 training sample; whereas the images generated by CCS-GAN are qualitatively reasonable and constitute a substantial improvement, although they remain less than clinically accurate.  It can be seen that basic Cycle-GAN at such low training volumes attempts to replicate bones and non-lung features.  Furthermore, basic Cycle-GAN appears to have unreasonably high contrast exaggerating the radiodensity of bones, while unreasonably suppressing the radiodensity of non-bone anatomy.  Finally, basic Cycle-GAN is only able to synthesize few if any pulmonary features.  Conversely CCS-GAN is able to synthesize relatively superior pulmonary features. With 50 COVID-10 images, it appears to generate ground glass opacities.  As such, it is clear that CCS-GAN, as it includes pulmonary segmentation, is vastly superior to a basic cycle-GAN in this qualitative comparison, as the basic cycle-GAN does not yield reasonable image quality at these low training volumes.

Quantitative analysis using these generated images was performed using AlexNet and VGG19; however, the results are not shown as baseline models achieve very poor performance, as anticipated because the synthetic images using only part of the CCS-GAN methodology are of unacceptable quality.  As such, we conclude from these qualitative ablation experiments that the entire CCS-GAN approach is necessary to achieve the reported qualitative results and quantitative performance.

\subsection{Conclusion and Future Work}
Diagnostic classification using few positive training example images is an important problem, because in the early stages of a pandemic, there may be a substantial lag between when the disease is of global health concern and when large datasets are publicly available.  A similar problem may also be encountered if screening for rare diseases in which few example images have been collected. 
A novel methodology is presented that combines intensity based pixel wise segmentation with Cycle Consistent Generative Adversarial Networks to generate synthetic COVID-19 CT scans with one order of magnitude fewer positive training examples than have been previously demonstrated. 
As such, CCS-GAN allows CNN classifiers to achieve high performance in COVID-19 diagnositc classification from the CT modality 
using only 10 CT-scan slices from 10 positive cases. 
As future work, we wish to extend this approach to be able to classify with only a single example of the target disease.  Furthermore, in future work we wish to determine if the reported results are only valid for detection of COVID-19 or if the CCS-GAN approach is capabile of screening the presence of other pulmonary diseases with few examples.  We remain optimistic that in the near future, the need for very large training datasets of a target disease will no longer be a hindrance to the timely development of accurate AI-based screening algorithms.


\end{document}